%% file: main.tex
\documentclass[sigconf, nonacm]{acmart}

\makeatletter

\usepackage[]{hyperref}
\usepackage{pifont}
\newcommand{\cmark}{\ding{51}}%
\newcommand{\xmark}{\ding{55}}%
\usepackage{multirow}
\usepackage{balance}
\usepackage{listings}
\usepackage{float} 

\begin{document}

\title{Stream-HLS: Towards Automatic Dataflow Acceleration}

\author{Suhail Basalama}
\email{basalama@ucla.edu}
\orcid{0000-0002-8301-8411}
\affiliation{%
  \institution{University of California, Los Angeles}
  \city{Los Angeles}
  \state{California}
  \country{United States}
}

\author{Jason Cong}
\email{cong@cs.ucla.edu}
\orcid{0000-0003-2887-6963}
\affiliation{%
  \institution{University of California, Los Angeles}
  \city{Los Angeles}
  \state{California}
  \country{United States}
 }

\input{sections/abstract}




\keywords{FPGA; HLS; MLIR; Dataflow Architecture; MINLP; Streaming}

\maketitle

\vspace{3in}

\input{sections/introduction}

\input{sections/motivation}

\input{sections/methodology}

\input{sections/implementation}

\input{sections/evaluation}

\input{sections/related_work}
\input{sections/conclusion}

\bibliographystyle{ACM-Reference-Format}
\balance
\bibliography{references}

\end{document}

%% file: sections/abstract.tex
\begin{abstract}
High-level synthesis (HLS) has enabled the rapid development of custom hardware circuits for many software applications. However, developing high-performance hardware circuits using HLS is still a non-trivial task requiring expertise in hardware design. Further, the hardware design space, especially for multi-kernel applications, grows exponentially. Therefore, several HLS automation and abstraction frameworks have been proposed recently, but many issues remain unresolved. These issues include: 1) relying mainly on hardware directives (pragmas) to apply hardware optimizations without exploring loop scheduling opportunities. 2) targeting single-kernel applications only. 3) lacking automatic and/or global design space exploration. 4) missing critical hardware optimizations, such as graph-level pipelining for multi-kernel applications.

To address these challenges, we propose a novel methodology and framework on top of the popular multi-level intermediate representation (MLIR) infrastructure called Stream-HLS. Our framework takes a C/C++ or PyTorch software code and automatically generates an optimized dataflow architecture along with host code for field-programmable gate arrays (FPGAs). To achieve this, we developed an accurate analytical performance model for global scheduling and optimization of dataflow architectures. Stream-HLS is evaluated using various standard HLS benchmarks and real-world benchmarks from transformer models, convolution neural networks, and multilayer perceptrons. Stream-HLS designs outperform the designs of prior state-of-the-art automation frameworks and manually-optimized designs of abstraction frameworks by up to $79.43\times$ and $10.62\times$ geometric means respectively. Finally, the Stream-HLS framework is modularized, extensible, and open-sourced at \url{https://github.com/UCLA-VAST/Stream-HLS} (\url{https://doi.org/10.5281/zenodo.14585909}).

\end{abstract}

%% file: sections/introduction.tex
\section{Introduction}

With the slowdown of Moore's law~\cite{schaller1997moore} and the limits of Dennard's scaling~\cite{bohr200730, dennard2018perspective}, industry and academia have turned to specialized hardware architectures. Major tech companies developed their custom hardware accelerators as application-specific integrated circuits (ASICs) such as Google's TPUs~\cite{jouppi2017datacenter}, or field-programmable gate arrays (FPGAs) such as Microsoft cloud projects~\cite{chung2018serving, microsoft}.

\begin{figure}[h]
    \centering
    \includegraphics[width=1\linewidth]{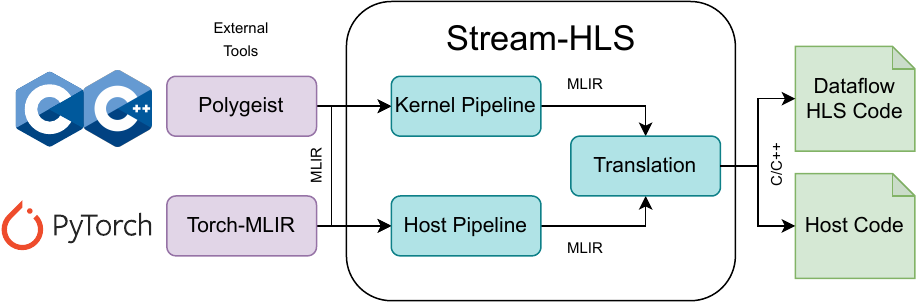}
    \caption{Stream-HLS Framework Overview}
    \label{fig:framework}
\end{figure}

While designing hardware accelerators with hardware description languages (HDLs) such as Verilog and VHDL attain the best speedups and energy efficiency, this process demands significant hardware expertise and can span several months to years. Fortunately, advancements in high-level synthesis (HLS) have elevated the abstraction level from HDLs to C/C++. As a result, HLS combined with the reconfigurability of FPGAs~\cite{cong2022fpga} has enabled them to compete with GPUs and CPUs in various domains such as machine learning~\cite{fracbnn, flexcnn, finn, caffeine}, image processing~\cite{dse_image_processing, clockwork}, genome sequencing~\cite{genome, lo2020algorithm}, and scientific computing applications~\cite{scicomp, muslim2017efficient}. Despite this democratization of FPGA programming, designing high-performance accelerators with HLS still requires searching a huge design space and expertise in hardware design. To address these challenges, many prior works have focused on developing automation and abstraction tools and methodologies.

\input{tables/prior_works}

A typical HLS flow for accelerating a software application requires multiple iterations of code refactoring and transformation, as well as the insertion of directives (pragmas) to implement specific hardware optimizations, such as loop pipelining and loop unrolling. Furthermore, many software applications consist of multiple smaller tasks, known as kernels, such as matrix multiplication or the layers of a deep neural network (DNN). This makes optimizing multi-kernel applications more challenging, requiring global analysis, scheduling, and resource allocation for different kernels. Table~\ref{tab:compare} summarizes previous attempts towards abstracting and automating hardware design through HLS.

Numerous previous studies proposed automation tools that explore different aspects of the vast design space of transforming software applications into hardware accelerators. AutoDSE~\cite{autodse} and HARP~\cite{harp}, for instance, are state-of-the-art design space exploration tools that find the best hardware optimization directives (pragmas) to insert into the software application. These tools rely on another compiler called Merlin~\cite{merlin} to perform code refactoring suitable for HLS tools, but without changing the original code schedule. However, pragma/directive insertion without loop scheduling may not be sufficient to achieve good quality results. Furthermore, this line of work focuses on small applications with a single or a small number of kernels, such as those in the Polybench~\cite{Polybench} suite.

Other works such as AutoSA~\cite{autosa} and SODA~\cite{soda}, introduced automated frameworks that convert C/C++ or domain-specific language (DSL) code into an optimized HLS code through both code transformation and pragma insertion. While these frameworks generate high-performance designs, their scope is limited to supporting only single-kernel applications (e.g., a single convolution layer). For multi-kernel applications, users need to manually integrate the different parts of the application. With the huge design space of global scheduling, this becomes a challenging task.

Other efforts, such as HeteroCL~\cite{heterocl}, and Allo~\cite{allo}, raise the level of abstraction for HLS from C/C++ to Python, allowing code transformation and directive optimizations within Python code. Although this approach enables the decoupling of algorithm design from the hardware optimization process at a higher level, it still necessitates the manual optimization of an expert to navigate the design space to produce high-performance designs.

Lastly, a few prior works built automated frameworks that can take multi-kernel applications and convert them into optimized dataflow designs by performing code transformation and pragma insertion, such as ScaleHLS~\cite{scalehls}, HIDA~\cite{hida}, and POM~\cite{pom}. However, these frameworks lack two important aspects: 1) Streaming capabilities. 2) Global performance modeling for streaming architectures. Our work, Stream-HLS, falls under this category but produces globally optimized dataflow designs with streaming capabilities.

In this study, we present an end-to-end framework that takes a sequential multi-kernel program written in C/C++ or PyTorch and automatically converts it into an optimized HLS code representing a dataflow architecture as Figures~\ref{fig:framework} and~\ref{fig:3mm_dag}b depict. While our methodology applies to FPGA and ASIC HLS tools, we use FPGAs for demonstration purposes. 

The main contributions of our work are as follows:
\begin{enumerate}
    \item an automatic, open-source, end-to-end framework to convert sequential multi-kernel applications into optimized parallel dataflow architectures with streaming capabilities;
    \item an MLIR~\cite{lattner2020mlir} library that performs multiple optimizations on a dataflow graph including dataflow canonicalization and the conversion of shared buffers to FIFOs when possible;
    \item an accurate performance model for dataflow architectures with shared-buffer and/or FIFO inter-task communications;
    \item a global coordinated loop scheduling approach considering node-level pipelining, graph-level pipelining, and node-level parallelization using mixed integer nonlinear programming (MINLP);
    \item a comprehensive evaluation against baseline models optimized by Vitis HLS~\cite{vitis} and other SOTA frameworks achieving geometric speedups between $5.42\times$ and $2270.25\times$.
\end{enumerate}

The paper starts with a motivating example (Section~\ref{sec:motivation}). Next, we present our methodology, detailing the process of converting sequential designs into parallelized dataflow architectures as well as our novel performance model and scheduling approach (Section~\ref{sec:method}). We then discuss the implementation of this methodology as an automated framework on top of MLIR (Section~\ref{sec:implementation}). Finally, we provide a comprehensive evaluation on a variety of benchmarks demonstrating the optimizations of Stream-HLS (Sections~\ref{sec:eval}).

%% file: tables/prior_works.tex
\begin{table*}[ht]\centering
\resizebox{\textwidth}{!}{
\begin{tabular}{l|c|c|c|c|c|c|c}
\hline

\multicolumn{1}{c|}{\multirow{2}{*}{\textbf{Framework (Venue)}}} &  
\multirow{2}{*}{\textbf{Front-end}} & 
\multirow{2}{*}{\textbf{App. Domain}} & 
\textbf{Multi-kernel} & 
\textbf{Loop} & 
\textbf{Graph Pipelining} & 
\textbf{Adaptive} 
&\textbf{Performance}
\\

& 
& 
& 
\textbf{Apps.} & 
\textbf{Pipelining} & 
\textbf{(Streaming Dataflow)} & 
\textbf{Parallelization} 
&\textbf{Model}
\\ 

\hline\hline
AutoDSE~\cite{autodse}(TODAES'22)            &          C/C++ &      General & \textcolor{orange}{limited} & \textcolor{green}{\cmark}  & \textcolor{red}{\xmark}    & \textcolor{red}{\xmark}         & HLS Report \\
HARP~\cite{harp}(ICCAD'23)             &          C/C++ &      General & \textcolor{orange}{limited} & \textcolor{green}{\cmark}  & \textcolor{red}{\xmark}    & \textcolor{red}{\xmark}         & ML-based \\
SODA~\cite{soda}(ICCAD'18)             &            DSL &     Stencils & \textcolor{red}{\xmark}     & \textcolor{green}{\cmark}  & \textcolor{green}{\cmark}  & \textcolor{green}{\cmark}       & Analytical \\
AutoSA~\cite{autosa}(FPGA'21)              &          C/C++ &       Affine & \textcolor{red}{\xmark}     & \textcolor{green}{\cmark}  & \textcolor{green}{\cmark}  & \textcolor{green}{\cmark}       & Analytical \\
HeteroCL~\cite{heterocl}(FPGA'19)              & Python/PyTorch &      General & \textcolor{green}{\cmark}   & \textcolor{orange}{Manual} & \textcolor{red}{\xmark}    & \textcolor{orange}{Manual}      & N/A \\
Allo~\cite{allo}(PLDI'24)              & Python/PyTorch &       Affine & \textcolor{green}{\cmark}   & \textcolor{orange}{Manual} & \textcolor{orange}{Manual} & \textcolor{orange}{Manual}      & N/A \\
ScaleHLS~\cite{scalehls}(HPCA'22)              &  C/C++/PyTorch &       Affine & \textcolor{green}{\cmark}   & \textcolor{green}{\cmark}  & \textcolor{red}{\xmark}    & \textcolor{orange}{C/C++ only}  & Analytical    \\
HIDA~\cite{hida}(ASPLOS'24)            &  C/C++/PyTorch &       Affine & \textcolor{green}{\cmark}   & \textcolor{green}{\cmark}  & \textcolor{red}{\xmark}    & \textcolor{orange}{C/C++ only}   & Analytical   \\
POM~\cite{pom}(HPCA'24)            & DSL            &       Affine & \textcolor{green}{\cmark}   & \textcolor{green}{\cmark}  & \textcolor{red}{\xmark}    & \textcolor{green}{\cmark}       & Analytical \\
\textbf{Stream-HLS (this work)}           &  C/C++/PyTorch &       Affine & \textcolor{green}{\cmark}   & \textcolor{green}{\cmark}  & \textcolor{green}{\cmark}  & \textcolor{green}{\cmark}   & Analytical  \\

\hline
\end{tabular}
}
\caption{Stream-HLS Comparison with Prior Frameworks}
\label{tab:compare}
\vspace{-0.25in}
\end{table*}

%% file: sections/motivation.tex
\section{Motivation}
\label{sec:motivation}

\begin{figure}[h]
    \centering
    \includegraphics[width=1\linewidth]{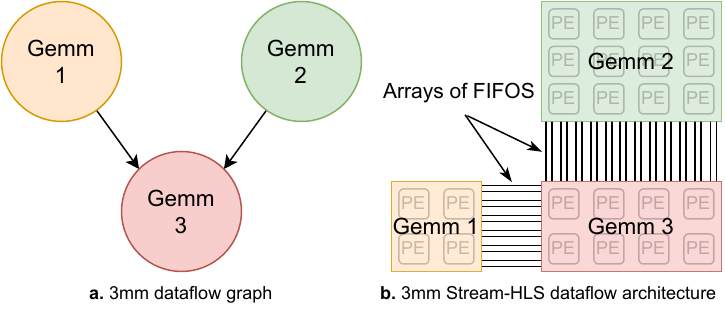}
    \caption{Dataflow Graph and Stream-HLS Architecture of \textit{\textbf{3mm}}~\cite{Polybench}}
    \label{fig:3mm_dag}
\vspace{-0.1in}
\end{figure}

Let us take a look at a simple multi-kernel application from the Polybench suite~\cite{Polybench} called \textit{\textbf{3mm}}. This application is composed of three matrix multiplications (gemm) where the outputs of the first two are used as input to the third matrix multiplication. Figure~\ref{fig:3mm_dag}a shows the high-level dataflow graph for this benchmark. Given such a graph, the challenge we are tackling is how to automatically generate an optimized accelerator under hardware constraints.

\begin{figure*}[t]
    \centering
    \includegraphics[width=1\linewidth]{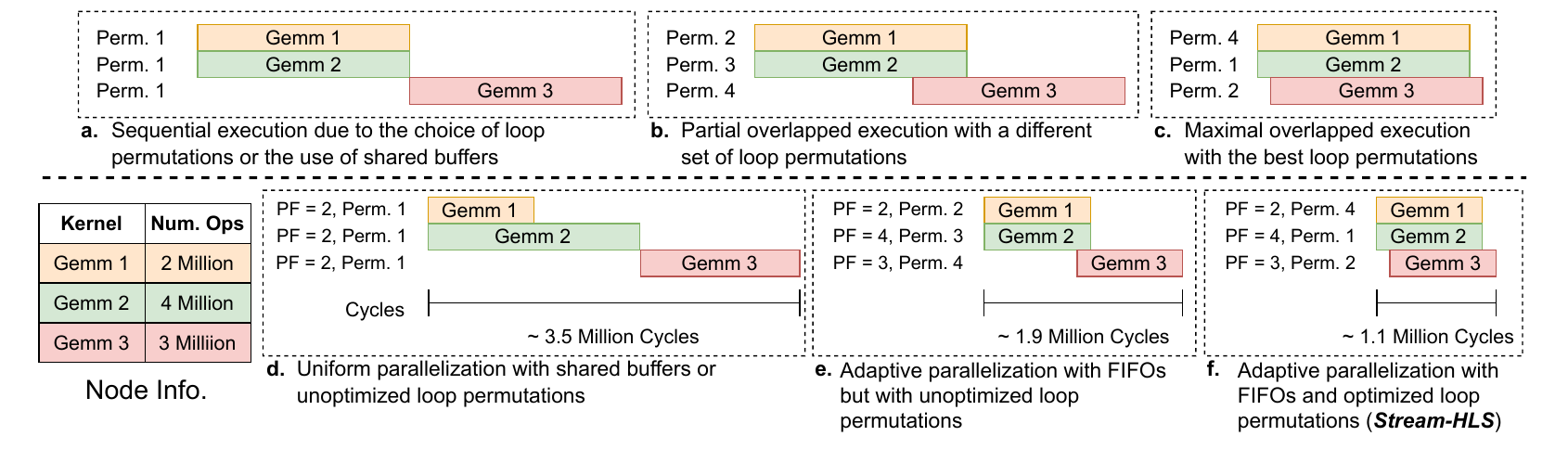}
\vspace{-0.3in}
    \caption{Execution \textit{\textbf{3mm}} Execution Traces under Different Loop Permutations and Parallelization Factors}
    \label{fig:3mm_traces}
\vspace{-0.15in}
\end{figure*}

We identified three critical but often conflicting optimizations that determine the overall performance. The design space for these optimizations, even for this simple example, is massive.

\subsection{Node-level Pipelining} 

Loop pipelining is one of the most important optimizations in HLS. 
Ideally, with enough resources, a loop can be pipelined with an initiation interval (II) of 1; however, loop-carried dependencies (true data dependencies between write and read operations of consecutive iterations) may prohibit achieving an II of 1. One solution to this is to perform loop permutation that increases the dependence distance achieving an II of 1. For a \textit{\textbf{gemm}} node (perfectly-nested loops of depth 3), there are $3!=6$ permutations, four of which lead to an II of 1, and two (where the reduction loop is innermost) lead to an II greater than 1. Prior works such as ScaleHLS~\cite{scalehls} and HIDA~\cite{hida} address this issue by automatically permuting reduction loops to be the outermost for all nodes of an application dataflow graph, thus maximizing the loop-carried dependence distances. However, would this local optimization (i.e. for each node regardless of the whole dataflow graph) be always beneficial? This takes us to the next important optimization, namely, graph-level pipelining.

\subsection{Graph-level Pipelining}

Edges in a dataflow graph represent true data dependencies, and communication between nodes can occur via shared buffers or FIFO streaming channels. Using shared buffers requires dependent nodes to execute completely sequentially to avoid any dependency violations, which limit graph-level pipelining (Figure~\ref{fig:3mm_traces}a). In contrast, streaming channels can be used to forward data elements that are ready before a node finishes its execution, enabling overlapped execution of dependent nodes as depicted in Figures~\ref{fig:3mm_traces}b and ~\ref{fig:3mm_traces}c.

The feasibility of using streaming channels depends on the data order forwarded between nodes, which in turn is determined by the loop permutation of each node. Therefore, assessing the potential for overlapped execution requires examining the loop permutations across the entire graph. Figure~\ref{fig:3mm_traces}a illustrates the execution traces for \textit{\textbf{3mm}}, where nodes run sequentially, either due to mismatched data order or exclusive use of shared buffers, a common approach in prior work~\cite{autodse, harp, heterocl, scalehls, hida, pom}. Figure~\ref{fig:3mm_traces}b presents a scenario where certain loop permutations enable partial overlap between \textbf{\textit{Gemm 1}}, \textbf{\textit{Gemm 2}}, and \textbf{\textit{Gemm 3}} using streaming channels. Figure~\ref{fig:3mm_traces}c shows an even more efficient set of permutations.

In this case, there are $3!\times3!\times3!=216$ possible loop permutation combinations for this small graph, making it challenging for designers to find the optimal set that balances \textbf{node-level} and \textbf{graph-level pipelining}. Additionally, some commercial HLS tools, like Vitis HLS~\cite{vitis}, may provide overly optimistic estimates for dataflow designs, which do not always reflect actual hardware performance. For example, Vitis HLS reports the same performance for designs in Figures \ref{fig:3mm_traces}b and \ref{fig:3mm_traces}c, but RTL simulations reveal the differences in their execution times.

\subsection{Node-level Parallelization}

The third optimization is parallelization within individual nodes of a dataflow graph. For high-performance accelerators, maximizing the use of computation resources, particularly DSPs (digital signal processing components), is essential. This can be achieved through complete loop unrolling, or partial loop unrolling using the divisors of loops' trip counts as done in ~\cite{hida, scalehls, autodse, harp}. We call the product of the unroll factors of a node's loop nest the parallelization factor (\textit{\textbf{PF}}). In single-kernel applications, parallelism is simply maximized by maximizing the kernel's \textit{\textbf{PF}}. However, in multi-kernel applications, hardware resources must be distributed across the computation nodes in the dataflow graph. Using a single \textit{\textbf{PF}} for all nodes in a graph can lead to imbalanced execution and under-utilization of hardware resources especially for imbalanced workloads, a scenario we call \textit{\textbf{uniform parallelization}}. For example, in the \textit{\textbf{3mm}} example shown in Figure~\ref{fig:3mm_traces}, a uniform \textit{\textbf{PF}} of 2 results in a performance of around 3.5 million cycles (Figure~\ref{fig:3mm_traces}d).

An alternative approach is \textit{\textbf{adaptive parallelization}}, which assigns PFs proportional to each node's workload (Figures~\ref{fig:3mm_traces}e and ~\ref{fig:3mm_traces}f). However, we found that adaptive parallelization alone is not always optimal because loop permutations still affect graph-level pipelining. For instance, even with the same adaptive parallelization, the design in Figure~\ref{fig:3mm_traces}f outperforms the one in Figure~\ref{fig:3mm_traces}e due to changing loop permutations.

Abstraction frameworks like HeteroCL~\cite{heterocl}, and Allo~\cite{allo} require users to manually specify unroll factors for each node, which is challenging especially for multi-kernel applications. In the \textit{\textbf{3mm}} example, there are 5 parameters representing the problem size. For the \textit{\textbf{3mm}} medium case in the Polybench suite~\cite{Polybench}, these parameters are $\{180, 190, 200, 210, 220\}$, with $\{18, 8, 12, 16, 12\}$ divisors, yielding a Cartesian product of 331,776 unique designs.

\subsection{Design Space Size}
The total design space, for this simple example, under these three optimizations, becomes $216\times331,776=71,663,616$ possible designs! Given only these three optimizations, two challenges arise: 
\begin{itemize}
    \item \textbf{\textit{How to estimate the performance of a dataflow design point with a mixture of shared-buffer and FIFO communication interfaces?}}
    \item \textbf{\textit{How to search this huge design space?}}
\end{itemize}
Stream-HLS addresses these two challenges by developing an accurate performance model capturing the three optimizations (including shared buffer and streaming interfaces) and formulates the design space exploration problem as solving MINLPs.

%% file: sections/methodology.tex
\section{Methodology}
\label{sec:method}

\subsection{Definitions and Preliminaries}

\textbf{Affine kernel} is any computation algorithm consisting of arithmetic statements enclosed by nested loops where loop bounds and memory accesses are affine functions of the loop iterators~\cite{pluto}.

\noindent
\textbf{Perfectly nested loops} are loops that are entirely contained within one another without any intervening code, where only the innermost loop contains some statements.

\noindent
\textbf{Legal conversion of shared buffers to FIFOs} requires two necessary conditions:
\renewcommand{\labelenumi}{\textbf{\textit{Cond. \arabic{enumi}.}}}
\setlength{\leftmargini}{0.5in}
\begin{enumerate}
    \item \label{conds:cond1} The number of writes to a shared buffer must equal the number of reads from the buffer.
    \item \label{conds:cond2} The order a producer task writes data to the buffer must match the order in which a consumer task reads the data.
\end{enumerate}

\subsection{Input Program}
Stream-HLS takes as input any arbitrary affine program consisting of a list of perfectly nested loops with constant loop bounds operating on multi-dimensional arrays (scalars and $1D$ arrays included). Arrays can have arbitrary affine references (reads or writes) in any loop nest. Many of the common targets for hardware acceleration are applications that can be represented as affine programs. This encompasses a wide variety of layers and kernels used in real-world DNNs and scientific computing applications such as data reshaping/transformation, matrix multiplication, dot product, convolution, pooling, activation layers (ReLU, GeLU, Sigmoid), softmax, batch normalization, piecewise add and multiply, to mention a few.

\begin{figure}[h]
    \centering
    \vspace{-0.1in}
    \includegraphics[width=0.8\linewidth]{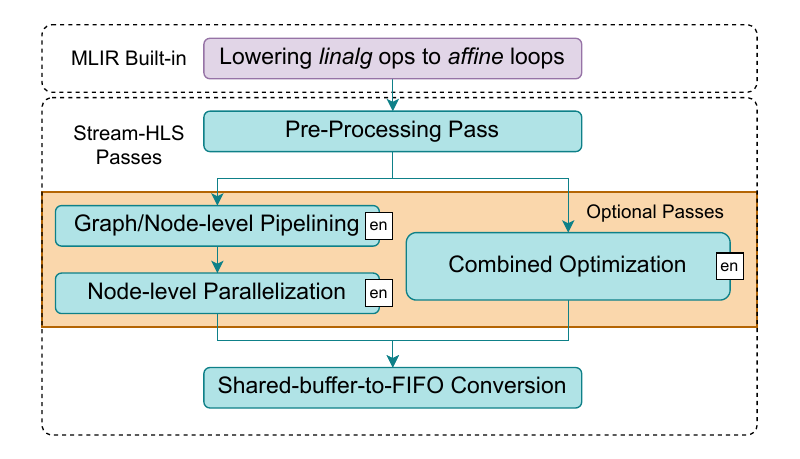}
    \vspace{-0.15in}
    \caption{Main Passes of Stream-HLS Kernel Pipeline}
    \label{fig:passes}
    \vspace{-0.25in}
\end{figure}
\subsection{Pre-Processing Pass}
\label{sec:preprocessing}
\subsubsection{\textbf{Dataflow Canonicalization}} 
\begin{figure}[h]
    \vspace{-0.1in}

    \centering
    \includegraphics[width=0.9\linewidth]{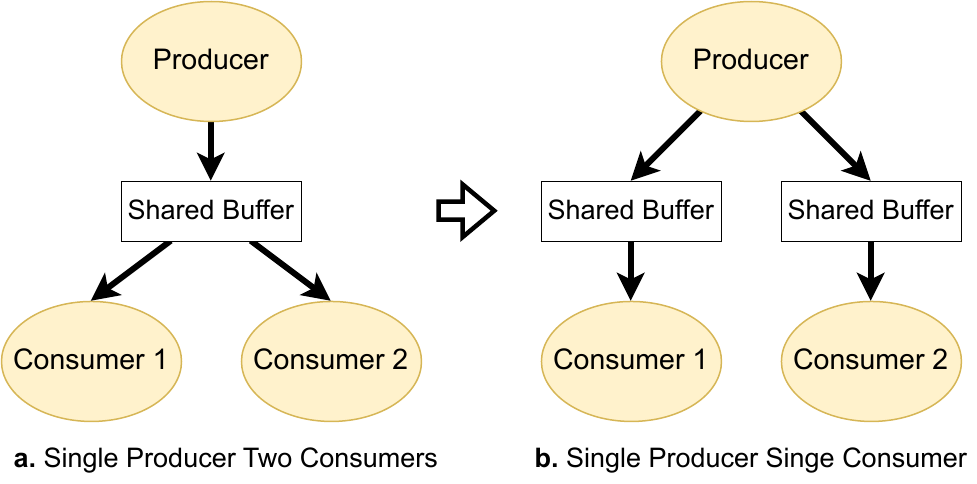}
    \caption{Dataflow Canonicalization}
    \label{fig:spsc}
    \vspace{-0.1in}
    
\end{figure}

The first step in our methodology is to make a program dataflow compatible. Some HLS tools impose rules a program must meet before it can be implemented as a canonical dataflow architecture. One rule is that all shared buffers must have a single-producer-single-consumer relationship. Thus, this step converts all single-producer-multi-consumer or multi-producer-multi-consumer shared buffers into single-producer-single-consumer shared buffers by duplicating the shared buffers. Figure~\ref{fig:spsc} illustrates this process.

\label{sec:fifo_conds}

\subsubsection{\textbf{Addressing Cond. ~\ref{conds:cond1}}}
\label{sec:cond1}

Many of the nodes of an application involve some data reuse in their input or output arrays. Let us take a simple example of matrix multiplication followed by matrix addition as shown in Listing~\ref{code:2mm}.

\begin{lstlisting}[label={code:2mm}, caption={Original Code for Matrix Multiplication Followed by Matrix Addition}, captionpos=b]
float C[32][32];
L1: for(int i = 0; i < 32; i++)
  L2: for(int j = 0; j < 32; j++)
    L3: for(int k = 0; k < 32; k++){
        float a = A[i][k];
        float b = B[k][j];
        float c = C[i][j];
        float tmp = c + a * b;
        C[i][j] = tmp;
    }
L4: for(int j = 0; j < 32; j++)
  L5: for(int i = 0; i < 32; i++)
    E[i][j] = C[i][j] + D[i][j];
\end{lstlisting}

Notice that array \texttt{C} is written to $32^3$ times in the first loop nest and is read only $32^2$ times in the second loop 
nest. In the first loop nest, the \texttt{C} array is reused locally 32 times by the innermost loop (commonly known as a reduction loop) with induction variable $k$. To satisfy \textit{\textbf{Cond. ~\ref{conds:cond1}}}, we transform the code such that only the final values (when $k = 31$) of the array \texttt{C} are written to a new buffer. Thus, in conjunction with the canonicalization step, we transform the code into the one shown in Listing~\ref{code:2mm1}.

\begin{figure}[ht] 
\begin{lstlisting}[label={code:2mm1}, caption={Transformed Version of Code in Listing~\ref{code:2mm}}, captionpos=b]
float C_shared_buff_1[32][32];
float C_shared_buff_2[32][32];
float C_local_buff[32][32];
L1: for(int i = 0; i < 32; i++)
 L2: for(int j = 0; j < 32; j++)
  L3: for(int k = 0; k < 32; k++){
   float a = A[i][k];
   float b = B[k][j];
   if(k==0)
    C_local_buff[i][j] = C_shared_buff_1[i][j];
   float c = C_local_buff[i][j];
   float tmp = c + a * b;
    C_local_buff[i][j] = tmp;
   if(k==31)
    C_shared_buff_2[i][j] = C_local_buff[i][j];
   }
L4: for(int j = 0; j < 32; j++)  
 L5: for(int i = 0; i < 32; i++)
  E[i][j] = C_shared_buff_2[i][j] + D[i][j];
\end{lstlisting}
\vspace{-0.2in}
\end{figure}

In this new equivalent code, array \texttt{C\_shared\_buff\_2} is written exactly $32^2$ times and read $32^2$ times, satisfying \textit{\textbf{Cond. ~\ref{conds:cond1}}}.

Dataflow canonicalization and addressing \textit{\textbf{Cond. ~\ref{conds:cond1}}} are performed jointly in the pre-processing pass in Figure~\ref{fig:passes}.

\subsection{Shared-buffer-to-FIFO Conversion}
\subsubsection{\textbf{Addressing Cond. ~\ref{conds:cond2}}}
\label{sec:cond2}
We utilize the concept of an access function ($AF$), which is a mapping from the loop induction variables to the indices of the array reference. For example, the write access function ($WAF$) of the store operation in line 15 of Listing ~\ref{code:2mm1} is $(i, j, k) \xrightarrow{} (i, j)$, and since the $k$ induction variable is not used, the access function simplifies to $(i, j) \xrightarrow{} (i, j)$. Similarly, the read access function ($RAF$) of the load operation (line 19) is $(j, i) \xrightarrow{} (i, j)$.

Now, we compare the write and read access functions. If they are equivalent, then the shared buffer can be safely replaced by a FIFO. In this example, since the functions are not equivalent, the shared buffer is retained. Nevertheless, we can see that permuting (interchanging) the $L1$ and $L2$ loops or the $L4$ and $L5$ loops makes the two access functions equivalent ($WAF = RAF$) and the shared buffer can be safely replaced with a FIFO. This brings us to the optional code transformation and performance modeling passes shown in the orange area in Figure~\ref{fig:passes}.

\subsection{Performance Modeling}
To perform any meaningful optimization, a performance model of dataflow architectures is needed.
\subsubsection{Dataflow Graph}
\begin{figure}[h]
    \centering
    \includegraphics[width=0.6\linewidth]{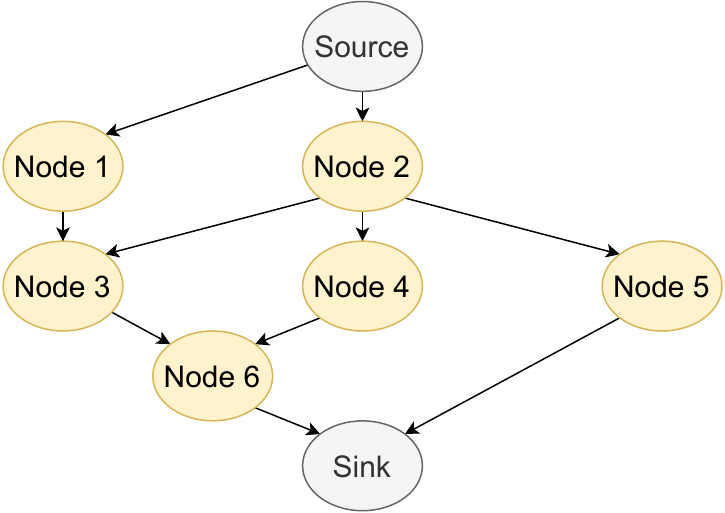}
    \caption{Dataflow Graph Example}
    \label{fig:dataflow}
    \vspace{-0.1in}
\end{figure}

In this step, a directed acyclic graph (DAG) is constructed to represent an application where:
\setlength{\leftmargini}{0.2in}    
\begin{itemize}
    \item \textbf{Node} represents a perfect loop nest with some statements representing a high-level operation such as matrix multiplication or DNN layers.
    \item \textbf{Edge} is a read-after-write (true/flow) dependency between nodes as Figure~\ref{fig:dataflow} depicts.
\end{itemize}

Nodes can have an arbitrary number of different inputs and a single output. The single output can then be an input to one or more other nodes. Each node $n$ contains critical information relevant to its performance in hardware (see Table~\ref{tab:node_info}). 

\begin{table}[h]
    \centering
    \begin{tabular}{c|l}
        \hline
        \textbf{Symbol} & \multicolumn{1}{c}{\textbf{Description}} \\ \hline \hline
        $II_n$ & Achievable initiation interval \\
        $U_n$ & Number of DSPs used for arithmetic operations \\
        $LR_n^{n'}$ & Relative$^*$ time of last read from input node $n'$ \\
        $RAF_n^{n'}$ & Read access function from input node $n'$ \\
        $FW_n$ & Relative$^*$ time of first write to the output \\
        $LW_n$ & Relative$^*$ time of last write to the output \\ 
        $WAF_n$ & Write access function to the output \\
        \hline
    \end{tabular}
\text{* relative to the absolute starting time $st(n)$ of the node}
    \caption{Node Information}
    \label{tab:node_info}
\vspace{-0.3in}
\end{table}

The achievable initiation interval ($II_n$) and DSP factor ($U_n$) are calculated based on characteristics of the loop nest and memory references, and the type of arithmetic operations it contains. 
The times of the last read ($LR_n^{n'}$), first write ($FW_n$), and last write ($LW_n$) are extracted from the order (permutation) of the loop nest and the achievable $II$. We define a time function $t(indices)$ that takes the indices (locations) of a cell of an array and returns a scalar number representing the time of that access. The function is calculated utilizing the order of the loops and their upper and lower bounds.
For example, let us consider the first loop nest in Listing~\ref{code:2mm1} to be node $n$. The time of the first and last writes to \texttt{C\_shared\_buff\_2} (line 15) are calculated using a time function as follows: \(t(i, j, k=31) = II_n(32^2i + 32j + k) \Rightarrow t(i, j) = II_n(32^2i + 32j + 31)\). Then, \(FW_n = t(0, 0) = 31\times II_n\) and \(LW_n = t(31, 31) = 32767\times II_n\). Note that these time equations depend on two things:
\setlength{\leftmargini}{0.2in}    
\begin{itemize}
    \item The permutations of the loops containing the memory access.
    \item The bounds of the loops containing the memory access.
\end{itemize}

This will be relevant for the performance model and the three optimizations we discussed in the motivation section.

\subsubsection{Performance Model Equations}
Next, we sort the DAG in a topological order, and for each node $n$, we calculate three equations as shown in Table~\ref{tab:performance_eqns}.

\begin{table}[h]
\vspace{-0.05in}

    \centering
    \begin{tabular}{l}
        \hline 
        \multicolumn{1}{c}{\textbf{Equation}} \\ \hline \hline
            \(st(n) = \max\limits_{n'\in \text{ins}(n)} \left(
            Arrives(n, n')
            \right)\) 
        \\ 
        where: \\
        \(Arrives(n, n') = 
        \begin{cases} 
                fw(n') & \text{if } WAF_{n'} = RAF_n^{n'}  \\
                lw(n') & \text{otherwise}
        \end{cases}\) \\ \hline

        \(fw(n) = st(n) + FW_n\) \\ \hline
        \(lw(n) = \max\limits_{n'\in ins(n)}\left(
            Depend(n, n') + Epilogue(n, n')
        \right)\) \\
        where: \\
        \(Depend(n, n') = \max(st(n) + LR_n^{n'}, lw(n'))\) \\
        \(Epilogue(n, n') = LW_n - LR_n^{n'}\) \\
        
        \hline
        
        $ins(n)$ is the set of input nodes of node $n$.
    \end{tabular}
    \caption{Performance Model Equations}
    \label{tab:performance_eqns}
\vspace{-0.25in}
\end{table}
\begin{figure}
    \vspace{-0.1in}
    \centering
    \includegraphics[width=\linewidth]{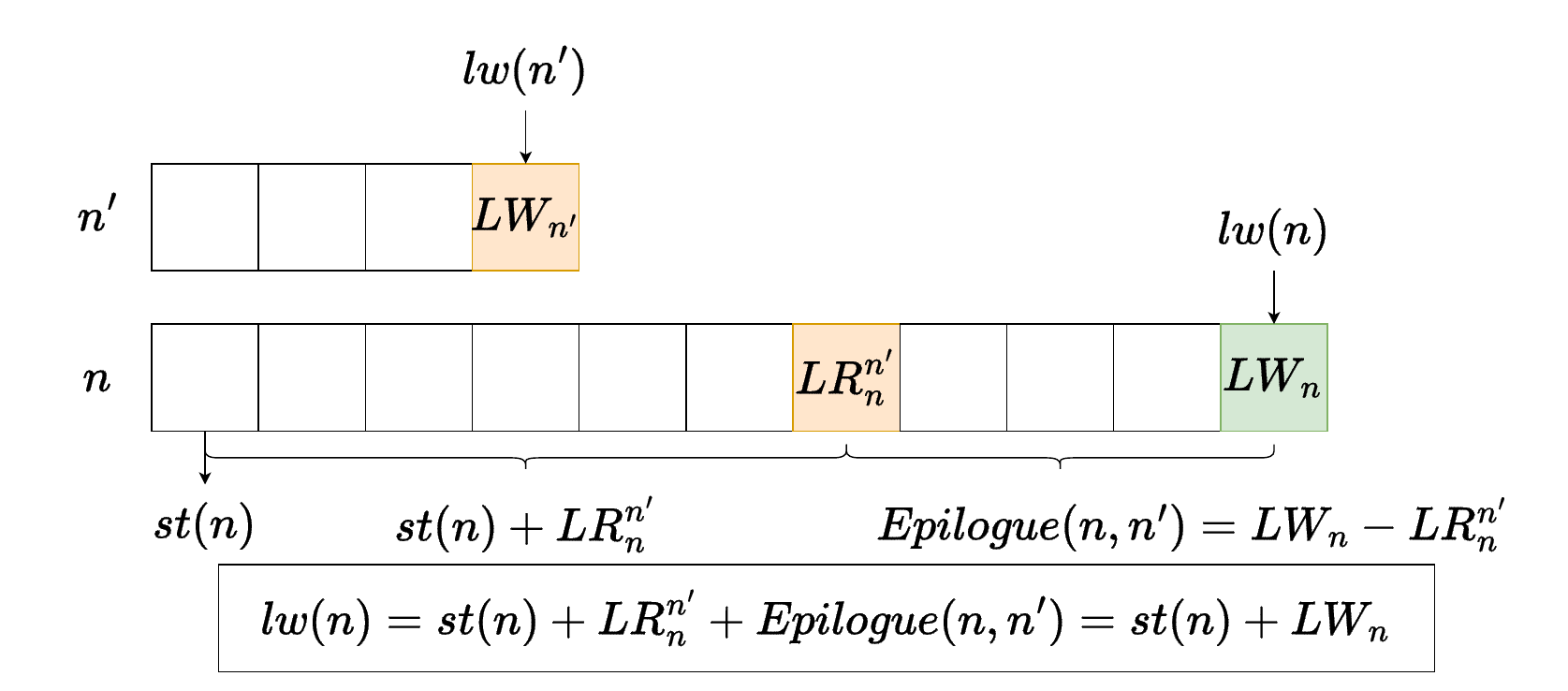}
    \caption{The First Case of the $lw(n)$: when a predecessor node $n'$ produces its data before node $n$ consumes it}
    \label{fig:depend_case1}
    \centering
    \includegraphics[width=\linewidth]{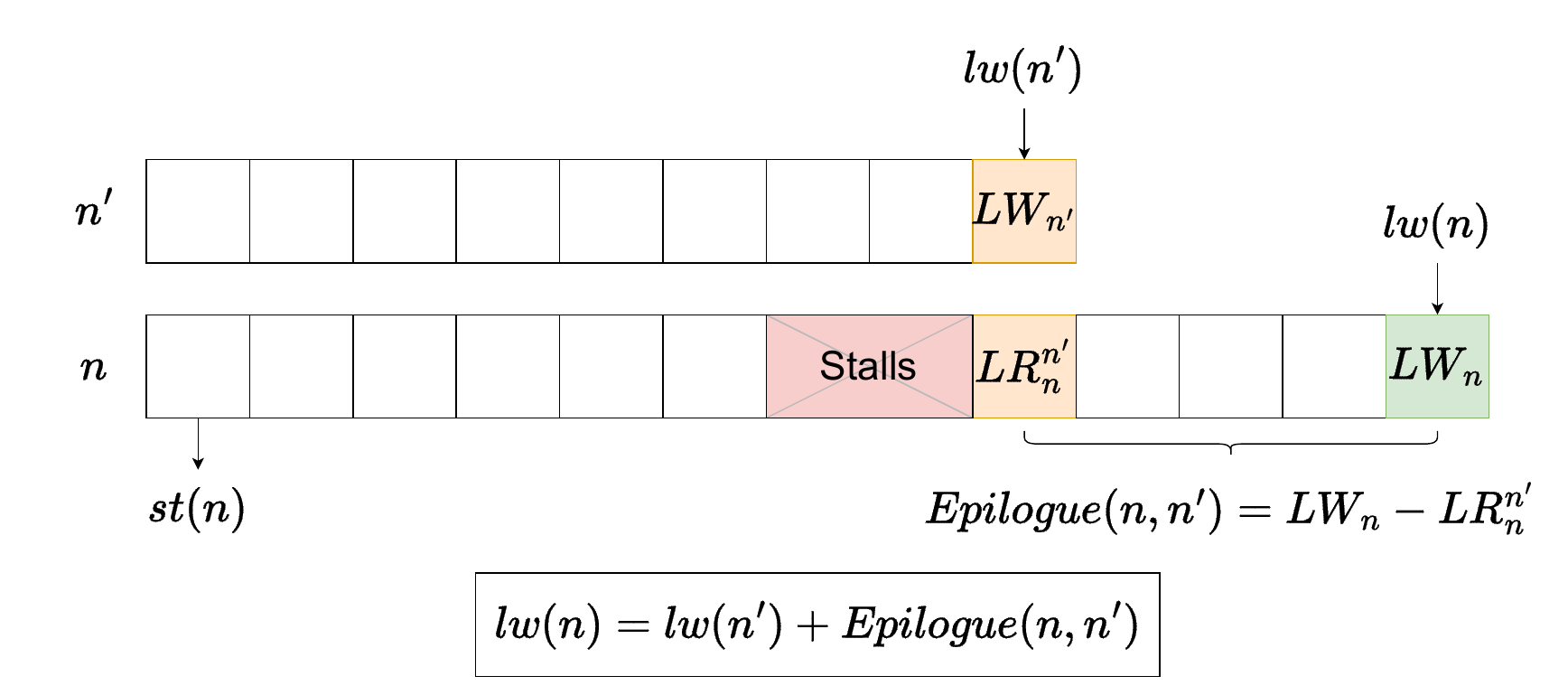}
    \caption{The Second Case of the $lw(n)$: when a consumer node $n$ must wait (stall) for data from its predecessor node $n'$}
    \label{fig:depend_case2}
    \vspace{-0.2in}
\end{figure}
\noindent$\bullet\ st(n)$ \textbf{represents the absolute starting time of node $n$.} A node cannot start execution until it has data on all its incoming edges. If a buffer can be replaced with a FIFO ($WAF_{n'} = RAF_n^{n'}$), then its arrival time is the absolute time of the first write by its predecessor node $n'$, otherwise (i.e., when a shared buffer cannot be replaced with a FIFO), the arrival time is the absolute end time of its predecessor node $n'$, which is equal to the last write $lw(n')$.

\noindent$\bullet\ fw(n)$ \textbf{represents the absolute time for the first write of node $n$}. It is calculated as the sum of the absolute starting time of the node ($st(n)$) and the relative first write time $FW_n$.

\noindent$\bullet\ lw(n)$ \textbf{represents the absolute time for the last write performed by node $n$, marking the time the node finishes its execution.} For each incoming node $n'$, we need to calculate two terms:

$\circ\ Epilogue(n,n')$ represents the remaining time after the last read by the current node $n$.

$\circ\ Depend(n, n')$ represents the time of the last read performed by node $n$, which is either the sum of the absolute starting time of $n$ and the relative time of its last write (when the write happens before the read as Figure~\ref{fig:depend_case1} depicts), or it is the absolute time of the last write performed by its predecessor node $n'$ plus the epilogue defined above since the last read of the current node cannot happen before the last write of the predecessor node as depicted in Figure~\ref{fig:depend_case2}.

The last write done by node $n$ (i.e., the end time of its execution) is then calculated as the maximum of the two terms above for all the incoming nodes. The total execution time of a dataflow graph is the time of the last write by the last node in the graph. For instance, the total execution time of the dataflow graph in Figure~\ref{fig:dataflow} is $lw(Sink)$. With a closed-form formula to estimate the performance of a design point (dataflow architecture), we can now optimize the code using the performance model as an objective function.

\subsection{Graph/Node-level Pipelining}
The achievable $II_n$ of each node and the degree of inter-node (graph-level) concurrency/pipelining is determined by the loop ordering of the loop nests of each node in the program. Therefore, for each node $n$ with $d$ nested loops in the dataflow graph, $d!$ binary indicator variables ($B_n$) are created to represent all loop permutations for that node. Note that different permutations of a node result in different values for the node information in Table~\ref{tab:node_info}. Thus, for each equation in the performance model, we augment the constants by multiplying them with the indicator variables as detailed in Table~\ref{tab:minlp_eqns}.

\begin{table}[t]
    \centering
    \begin{tabular}{l}
        \hline 
        \multicolumn{1}{c}{\textbf{Expanded Equation}} \\ \hline \hline
            \(st(n) = \max\limits_{n'\in \text{ins}(n)} \left(
                \sum\limits_{b \in B_n}
                \sum\limits_{b' \in B_{n'}}
                Arrives(n, n')\times b \times b'
                \right)
            \)
        \\ 
        where: \\
        \(Arrives(n, n') = 
        \begin{cases} 
                fw(n') & \text{if } WAF_{n'} = RAF_n^{n'}  \\
                lw(n') & \text{otherwise}
        \end{cases}\) \\ \hline

        \(fw(n) = st(n) + \sum\limits_{b \in B_n}FW_n \times b\) \\ \hline
        \(lw(n) = \max\limits_{n'\in ins(n)}\left(
            Depend(n, n') + Epilogue(n, n')
        \right)\) \\
        where: \\
        \(Depend(n, n') = \max(st(n) + \sum\limits_{b \in B_n}LR_n^{n'} \times b, lw(n'))\) \\
        \(Epilogue(n, n') = \sum\limits_{b \in B_n}(LW_n - LR_n^{n'})\times b\) \\
        
        \hline
        $ins(n)$ is the set of incoming nodes of node $n$. \\
        \( B_n \) is the set of binary indicator variables of node \( n \). 
    \end{tabular}
    \caption{Expanded Performance Model Equations}
    \label{tab:minlp_eqns}
\vspace{-0.3in}
\end{table}

For the starting time $st(n)$, we must consider the relationship between the producer and consumer nodes to determine whether a FIFO can be used or not (see the term $Arrives(n, n')\times b \times b'$) in Table~\ref{tab:minlp_eqns}. Finally, the MINLP program is constrained by making the sum of the indicator variables of each node equal 1 to choose one specific loop permutation. After solving the program, we perform a loop permutation on each node using the found solution. Equation~\ref{eq:minlp1} shows the objective function and constraints of the first MINLP.

\begin{equation}
\begin{aligned}
\min_{b \in B_n} \quad & lw(\text{Sink}) &  \forall n \in G\\
\textrm{s.t.} \quad & \sum_{b \in B_n} b = 1 & (\text{Perm. Const.}) \\
\end{aligned}
\label{eq:minlp1}
\vspace{-0.1in}
\end{equation}

\begin{figure}[h] 
\begin{lstlisting}[label={code:2mm_unroll}, caption={Tiled and Parallelized Version of Code in Listing~\ref{code:2mm1} with Array of FIFOs Communication}, captionpos=b]
L1_1: for(int i1 = 0; i1 < 32/Ti; i1++)
 L2_1: for(int j1 = 0; j1 < 32/Tj; j1++)
  L3_1: for(int k1 = 0; k1 < 32/Tk; k1++)
   #pragma HLS Pipeline
   L1_2: for(int i2 = 0; i2 < Ti; i2++)  //unroll
    L2_2: for(int j2 = 0; j2 < Tj; j2++) //unroll
     L3_2: for(int k2 = 0; k2 < Tk; k2++)//unroll
      //...omitted for brevity, i=i1*Ti+i2, j=j1*Tj+j2
      if(k1 == (32/Tk) - 1)
       fifos_2d[i2][j2].write(C_local_buff[i][j]);
L5_1: for(int i1 = 0; i1 < 32/Ti; i1++)
 L4_1: for(int j1 = 0; j1 < 32/Tj; j1++)
  #pragma HLS Pipeline
  L5_2: for(int i2 = 0; i2 < Ti; i2++)  //unroll
   L4_2: for(int j2 = 0; j2 < Tj; j2++) //unroll
    // i=i1*Ti+i2, j=j1*Tj+j2
    E[i][j] = fifos_2d[i2][j2].read() + D[i][j];
\end{lstlisting}
\vspace{-0.2in}
\end{figure}

\subsection{Node-level Parallelization}

So far, our analysis has been based on the granularity of single elements of data (scalars), so at a first glance, it seems our performance model does not consider the parallelization within each node. Nevertheless, if we think about the problem at the granularity of tiles of data, we can apply the same analyses on multi-dimensional tiles of data, rather than on scalars. To illustrate, let us look at the code in Listing~\ref{code:2mm_unroll}. It is a tiled version of the code shown in Listing~\ref{code:2mm1} (each loop is split into \texttt{Lx\_1}, and \texttt{Lx\_2}). Node-level parallelization is then achieved through unrolling the innermost loops and sending tiles of data between nodes, rather than single scalar values. Tiles of data are then communicated through partitioned shared buffers or using arrays of FIFOs (see Figure~\ref{fig:3mm_dag}b and lines 10 and 17 in Listing~\ref{code:2mm_unroll}). Another important aspect to consider is the tile size between dependent nodes. In order to keep an execution similar to the unparallelized code, we need to use the same tiling factors (\texttt{Ti} and \texttt{Tj}) between dependent nodes depicted in Listing~\ref{code:2mm_unroll}. Now, we can reason about a tiled program in a similar way to the original unparallelized one, but with variable loop bounds (\texttt{32/Ti},\texttt{32/Tj}, \texttt{32/Tk}). Therefore, the node information in Table~\ref{tab:node_info} become symbolic formulas in terms of the variable tiling factors. The values of these tiling factors (\texttt{Ti, Tj, Tk}) are to be found by another augmented performance model which is omitted for space. The new MINLP is shown in Equation~\ref{eq:minlp2}. 

\begin{figure}[h]
\centering
\begin{equation}
\begin{aligned}
    \min_{x \in X_n} \quad & lw(\text{Sink}) & \forall n \in G\\
    \textrm{s.t.} \quad & x = x' & (\text{Tile Size Const.}) \\
                        & ub \% x = 0 &  (\text{Divisor Const.}) \\
                        & \sum_{n \in G} ( U_n \prod_{x \in X_n} x ) \leq DSPs & (\text{DSP Const.}) \\
\end{aligned}
\label{eq:minlp2}
\end{equation}
\begin{flushleft}
Where: $x' \in X_{n'}, n' \in \text{ins}(n)$ \\
$x$ and $x'$ refer to the same array dimension. \\
$ub$ is the upper bound of the target loop.
\end{flushleft}
\vspace{-0.1in}
\end{figure}

For each node $n$ with $d$ nested loops in the dataflow graph, we create $d$ integer variables ($X_n$) to represent the tiling factors. The first constraint enforces the tile sizes between dependent nodes are equal, the second constraint enforces the tiling factors to be divisors of the loop bounds, and the last constraint ensures that the DSP utilization does not exceed the available DSPs of a given FPGA. $U_n$ is a factor that depends on the type of arithmetic operations and their hardware utilization.

\input{tables/error}

\subsection{Combined Optimization}
The previous two MINLPs can be used to find loop permutations and then tiling factors separately. The first MINLP explores the design space of node/graph-level pipelining, and the second MINLP explores the design space of node-level parallelization. However, since these two optimizations are interconnected, solving each MINLP separately may lead to sub-optimal solutions. Therefore, we create a third MINLP to explore the whole design space to capture both optimizations at once as shown in Equation~\ref{eq:minlp3}, which aims to find $d$ binary variables ($B_n$) and $d$ integer variables ($X_n$) representing loop permutation and tiling factors of a node $n$.

\begin{figure}[h]
\begin{equation}
\begin{aligned}
\min_{b \in B_n, x \in X_n} \quad & lw(\text{Sink}) & \forall n \in G\\
\textrm{s.t.} \quad & \sum_{b \in B_n} b = 1 & (\text{Perm. Const.}) \\
& x = x' & (\text{Tile Size Const.})\\
& ub \% x = 0 & (\text{Divisor Const.})\\
& \sum_{n \in G}{(U_n\prod_{x \in X_n}{x}) \leq DSPs} & (\text{DSP Const.})\\
\end{aligned}
\label{eq:minlp3}
\end{equation}
\begin{flushleft}
Where: same as Equation ~\ref{eq:minlp2} \\
\end{flushleft}
\vspace{-0.2in}
\end{figure}

%% file: tables/error.tex
\begin{table*}[t]
\resizebox{\textwidth}{!}{
    \centering
    \begin{tabular}{c|c|c|c|c|c|c}
    \hline
    \multirow{2}{*}{\textbf{Application}} & \multicolumn{3}{c|}{\textbf{Opt1} (Shared-buffers to FIFOs only)}  & \multicolumn{3}{c}{\textbf{Opt5} (Combined Optimization)}  \\ \cline{2-7}
                     & \textbf{RTL Cycles} & \textbf{Stream-HLS Cycles} & \textbf{Vitis Cycles} & \textbf{RTL Cycles} & \textbf{Stream-HLS Cycles} & \textbf{Vitis Cycles} \\
    \hline \hline

Autoencoder               & 2.98E+07        & \textbf{2.68E+07 ($0.90\times$)} & 2.57E+07 ($0.86\times$)          & 3.92E+04        & \textbf{3.62E+04 ($0.92\times$)} & 3.31E+04 ($0.84\times$)          \\
Residual MLP              & 1.84E+07        & \textbf{1.84E+07 ($1.00\times$)} & 1.68E+07 ($0.91\times$)          & 3.98E+04        & \textbf{3.31E+04 ($0.83\times$)} & 3.30E+04 ($0.83\times$)          \\ \hline
Residual Block            & 5.79E+07        & \textbf{5.79E+07 ($1.00\times$)} & 2.91E+07 ($0.50\times$)          & 2.09E+06        & 1.05E+06 ($0.50\times$)          & \textbf{1.83E+06 ($0.88\times$)} \\
DWSConv. Block            & 6.63E+06        & \textbf{6.53E+06 ($0.98\times$)} & 3.82E+06 ($0.58\times$)          & 1.35E+05        & 8.42E+04 ($0.63\times$)          & \textbf{1.34E+05 ($1.00\times$)} \\ \hline
Feed Forward              & 6.73E+07        & \textbf{6.73E+07 ($1.00\times$)} & 6.71E+07 ($1.00\times$)          & 6.60E+04        & \textbf{6.58E+04 ($1.00\times$)} & 6.56E+04 ($0.99\times$)          \\
Multi-Head Self Attention & 1.05E+07        & 1.06E+07 ($1.00\times$)          & \textbf{1.05E+07 ($1.00\times$)} & 3.51E+04        & \textbf{3.51E+04 ($1.00\times$)} & 3.47E+04 ($0.99\times$)          \\ \hline
3mm                       & 6.38E+07        & 6.38E+07 ($1.00\times$)          & \textbf{6.38E+07 ($1.00\times$)} & 4.91E+04        & \textbf{4.90E+04 ($1.00\times$)} & 4.80E+04 ($0.98\times$)          \\
atax                      & 1.28E+06        & \textbf{1.28E+06 ($1.00\times$)} & 8.00E+05 ($0.63\times$)          & 2.18E+03        & \textbf{2.10E+03 ($0.97\times$)} & 1.14E+03 ($0.52\times$)          \\ \hline
7mm Balanced              & 1.15E+06        & \textbf{1.15E+06 ($1.00\times$)} & 1.05E+06 ($0.92\times$)          & 5.68E+03        & \textbf{5.54E+03 ($0.97\times$)} & 4.24E+03 ($0.75\times$)          \\
7mm Imbalanced            & 4.29E+06        & \textbf{4.24E+06 ($0.99\times$)} & 4.19E+06 ($0.98\times$)          & 1.00E+04        & \textbf{9.20E+03 ($0.92\times$)} & 8.35E+03 ($0.83\times$)          \\ \hline
Geo. Mean                 &                 & $0.99\times$                     & $0.81\times$                     &                 & $0.85\times$                     & $0.85\times$                     \\
\hline 

    \end{tabular}
}
\caption{Stream-HLS Performance Model Prediction vs. Vitis HLS Prediction Compared to RTL Simulation Cycle Count}
\label{tab:errors}
\vspace{-0.3in}
\end{table*}

%% file: sections/implementation.tex
\section{Implementation}
\label{sec:implementation}

\subsection{MLIR Background}
MLIR (multi-level intermediate representation)~\cite{lattner2020mlir, lattner2021mlir} is a subproject of the famous LLVM compiler infrastructure~\cite{lattner2004llvm} that has been gaining popularity in the industry and academia. The novel idea of MLIR is the concept of \textit{dialects}, which is a collection of operations, types, and attributes to represent semantics at a specific level of abstraction, thus the name multi-level intermediate representation. MLIR includes many built-in dialects, such as the \textit{linalg} dialect that represents linear algebra operations like matrix multiplication and convolution. Another useful built-in dialect is the \textit{affine} dialect suitable for affine kernels, which enables efficient and reliable dependence analysis and loop transformations.

\subsection{External Tools}
To find solutions to our scheduling mathematical programs, we use the AMPL~\cite{ampl} interface with the Gurobi solver\cite{gurobi}. Further, Stream-HLS utilizes two external tools as front-ends for translating code into MLIR: 1) \textbf{Torch-MLIR}~\cite{torchmlir} for translating a PyTorch model into \textit{linalg} MLIR, and 2) \textbf{Polygeist}~\cite{polygeist} for translating C/C++ into \textit{affine} MLIR (see Figure~\ref{fig:framework}).

\subsection{Stream-HLS Implementation}
Stream-HLS mainly utilizes four built-in dialects: \textit{linalg}, \textit{affine}, \textit{memref}, \textit{arith}. Stream-HLS also adopts an updated version of an HLS dialect developed by prior works~\cite{scalehls, hida} for HLS-specific operations. In addition, for node-level parallelization, we created new types and operations to represent multi-dimensional array of streams/FIFOs that can be referenced using affine maps. The Stream-HLS library is written in C++ with more than 40K lines of code and is modularized enabling further development and extensions in the future. Stream-HLS has three main components described below.
\subsubsection{Host Pipeline}
While the MLIR infrastructure always ensures the validity of the IR at any point of the compiler passes, it is the responsibility of the developer to maintain the semantic equivalence of the program. A verification pipeline composed of compiler passes has been developed for this purpose. The pipeline takes the input program's IR (and model parameters for DNNs) and creates host code with a testbench comparing the outputs of the original software code with the outputs of the generated design.
\subsubsection{Kernel Pipeline}
The kernel pipeline consists of the critical conversion and scheduling passes implementing the ideas discussed in the previous section.

\subsubsection{Translation}
After transforming the original program into a host MLIR and kernel MLIR, the MLIRs are translated into C++ Vitis HLS code. The translation modules are based on the one developed by~\cite{scalehls, hida}. In addition to Vitis HLS, we developed translation for a SOTA HLS library called TAPA~\cite{tapa} built for dataflow designs.

\subsubsection{End-to-End Compilation}
Stream-HLS is fully automated and open-source. The framework takes a PyTorch model or C/C++ code as input, and with a push of a button, produces an optimized dataflow architecture C/C++ code along with host code that verifies the correctness of the implementation against the software golden results. Figure~\ref{fig:framework} illustrates a high-level overview of our framework.

%% file: sections/evaluation.tex
\section{Evaluation}
\label{sec:eval}

In this section, we validate the Stream-HLS performance model, evaluate Stream-HLS against Vitis HLS~\cite{vitis} baseline and prior works, and provide an in-depth evaluation of the different optimizations of Stream-HLS. All experiments are performed using Vitis HLS 2023.2~\cite{vitis} targeting AMD Alveo U280 Data Center Accelerator Card~\cite{u280} at 300MHz.
\subsection{Experimental Setup and Benchmarks}

We conducted experiments on five categories of applications: 1) seven kernels from the Polybench benchmark suite~\cite{Polybench}. 2) multi-head self attention and feed forward network found in transformer models~\cite{attention}. 3) A residual block~\cite{resnet} and a depthwise separable convolution block (DWSConv.)~\cite{mobilenet}(involving standard convolution, depthwise convolution, pointwise convolution, batch normalizaiton, and ReLU activation). 4) Two multilayer perceptron (MLP) networks: an encoder-decoder MLP~\cite{autoencoder}, and a 4-layer MLP with a residual connection. 5) Two synthetic 7 matrix multiplication kernels connected in series (7mm). For each application we conducted 5 experiments using different optimization levels of Stream-HLS as shown in Table~\ref{tab:opts_info}.
For all MINLPs, we set a timeout of 20 minutes and 2560 DSPs limit\footnote{Slightly less than the DSPs of a single super logic region (SLR) of the U280 FPGA} for the MINLP solver. All solutions were found in less than 2 minutes, except for \textbf{Opt4} and \textbf{Opt5} of the residual block and multi-head self attention block where they reached the 20 minute timeout due to the size of the design space. Due to the inaccuracies of Vitis HLS reports for dataflow designs, we conducted all experiments using RTL cycle-accurate simulation.
\input{tables/opts}

\input{tables/compare}

\subsection{Performance Model Validation}
To validate the performance model, we chose designs under \textbf{Opt1} and \textbf{Opt5}. \textbf{Opt1} designs are the baseline designs featuring a random level of graph/nodel-level pipelining determined by the original order of the loop nests with a mixture of shared buffers and FIFO interfaces. This can help us show the accuracy of our performance model under random configuration. Additionally, we show the predictions of our performance model for the fully-optimized designs (\textbf{Opt5}). For the randomized \textbf{Opt1} designs, Stream-HLS predictions are very close to the actual RTL simulation cycles while Vitis HLS estimates can be $0.50\times$ of the actual RTL simulation cycles. We suspect that such optimistic estimates of Viti HLS are due to the use of a simplified performance model for dataflow designs that uses the slowest node as the latency of the whole design. For \textbf{Opt5} designs, Stream-HLS predictions may not match the actual RTL simulation. The reason for this is that with node-level parallelization, Vitis HLS may not achieve the requested II for some nodes leading to a higher II than our performance model prediction. This is a difficult problem since the Vitis HLS compiler performs further code instrumentation that are hard to predict. In fact, with some manual code refactoring related to the coding style (without changing the schedule), Vitis HLS can achieve the expected II, but this is not automated yet in our framework and will be improved in the next version of Stream-HLS. Furthermore, we have noticed that the degradation in II is relative. For instance, if Stream-HLS predicts the II of a specific loop nest to be 1 for permutation 1 and 4 for permutation 2, then the achieved II by Vitis HLS for permutation 1 is lower (better) than the achieved II for permutation 4.

\subsection{Comparison with Prior Works}

In this subsection, we compare Stream-HLS with manually and automatically optimized designs generated by similar state-of-the-art frameworks. Allo~\cite{allo} and HeteroCL~\cite{heterocl} are abstraction frameworks that enable a user to optimize a schedule manually through schedule optimization primitives, such as loop reordering (permutation), loop piplining, and loop unrolling. What is unique about Allo compared to HeteroCL is the \textit{\textbf{compose}} primitive, which enables graph-level pipelining through streaming channels for multi-kernel applications. HIDA~\cite{hida}, ScaleHLS~\cite{scalehls} and POM~\cite{pom} are automatic hardware compilers that take a C/C++ or PyTorch model (or DSL for POM) and automatically generate HLS designs. However, these frameworks have two limitations: 1) They adopt shared buffers only for inter-task communication, which limits graph-level pipelining. 2) Automatic design space exploration is only available for the C/C++ and DSL front-ends, while the PyTorch front-end lacks DSE capabilities and requires a user to specify a single unroll factor to be applied for all loop nests (uniform parallelization).

For Allo and HeteroCL, we used designs manually optimized by the tools' authors targeting the same U280 FPGA. For HIDA, ScaleHLS, and POM we used their C/C++ and DSL front-ends to perform design space exploration to find the best designs targeting the U280 FPGA with the maximum 9024 DSP limit (whole board). For Vitis HLS~\cite{vitis} baseline designs, we did not perform any optimization except for the default pipelining applied by Vitis HLS. For Stream-HLS, we generated the designs using the combined optimization \textbf{Opt5} under three different DSP limit constraints 220, 2560 (single SLR), and 9024 DSPs. All designs finished RTL simulation except: 1) Allo's \textit{\textbf{3mm}} design, which ran into an infinite loop during the simulation, so we used cycles reported by Vitis HLS instead. 2) POM's \textbf{\textit{atax}} and \textit{\textbf{bicg}} designs, which failed during the DSE step with internal errors. Further, we also noticed that POM does not exceed the 220 DSP limit used in their paper despite setting the 9024 DSP limit in the configuration file, thus we added the 220 DSP limit experiments for Stream-HLS for a fair comparison.

Table~\ref{tab:comp} shows the absolute RTL cycles and the speedups of Stream-HLS designs over the prior frameworks. We can see that Stream-HLS designs are better or comparable to designs manually optimized by an expert in the case of Allo and HeteroCL at different DSP limits. The reason for this is that design space even for small applications can grow exponentially, making it difficult for a human experts to find the set of optimizations to apply. Conversly, Stream-HLS designs constantly outperform the designs of HIDA, ScaleHLS starting at the 1 SLR DSP limit and the designs of POM even at the 220 DSP limit. We believe this is due to the lack of graph-level pipelining and/or their performance modeling with the iterative DSE algorithms employed by these frameworks.
\input{tables/scalability}

Additionally, we compare the scalability of Stream-HLS MINLP DSE under the three DSP constraints with HIDA's~\cite{hida} (which uses the same DSE of ScaleHLS~\cite{scalehls}) and POM's~\cite{pom} iterative DSEs. We also compare the DSP percentage utilization of the three frameworks as depicted in Table~\ref{tab:scalability}. The runtime of Stream-HLS DSE outperforms HIDA's~\cite{hida} DSE with a geometric mean of $176.41\times$. POM's DSE runtime is similar to ours, but it produces inferior designs. In terms of DSP allocation, Stream-HLS constantly maximizes DSP utilization under the specified limit and problem size constraints thanks to our global scheduling approach. Conversely, under the same DSE configuration, HIDA and POM yield variable DSP utilization for different applications.

\input{tables/3mm}

\input{tables/ablation}

Lastly, Table~\ref{tab:3mm_perf} breaks down the latencies and DSP utilization of the three nodes of \textit{\textbf{3mm}}. We can see that under a specific DSP limit, Stream-HLS distributes the DSP resources among the three nodes while maximizing graph/node-level pipelining resulting in a total latency close to that of the slowest node. The DSEs of HIDA and POM result in a less even distribution of DSPs~\footnote{Total DSPs include top-level DSPs that are shared/reused between Gemm2 and Gemm3} and a sequential execution between \textbf{\textit{Gemm 1}} or \textbf{\textit{Gemm 2}} and \textbf{\textit{Gemm 3}} due to the lack of graph-level pipelining confirming our motivation example in Figure~\ref{fig:3mm_traces}. 

\subsection{Different Optimizations of Stream-HLS}

To isolate the impact of different Stream-HLS optimizations, we generate five designs for each application starting with only replacing shared buffers with FIFOs as a baseline (refer to Table~\ref{tab:opts_info}). We can see that optimizing loop order only (\textbf{Opt2}) can help maximize graph/node-level pipelining, achieving between $3.91\times$ and $7.28\times$ speedups. Next, applying node-level parallelization without considering loop order (\textbf{Opt3}) leads to speedups between $7.99\times$ and $63.85\times$. \textbf{Opt4} first optimizes loop order to maximize graph/node-level pipelining and then optimizes node-level parallelization in two separate MINLPs. This optimization yields even better performance, ranging between $27.7\times$ and $1300.42\times$. Finally, \textbf{Opt5}, which combine both graph/node-level pipelining and node-level parallelization in a single MINLP results in the best performance, with speedups ranging between $27.67\times$ and $1300.42\times$ with a geometric mean of $314.89\times$ compared to $152.30\times$ for \textbf{Opt4} designs. As expected, in some cases solving the two MINLPs can lead to sub-optimal designs due to the conflict between the graph/node-level pipelining and node-level parallelization optimizaitons as shown in the MLP and transformer designs. Based on our observations, we noticed that solving the two MINLPs separately usually leads to the same solutions found by the combined approach when the workloads are balanced, i.e., the number of computations per node are similar. Thus, we created two synthetic applications one with balanced nodes (7mm Balanced), and the second one with imbalanced nodes (7mm Imbalanced), and the results seem to confirm this.

%% file: tables/opts.tex
\begin{table}[h]
\resizebox{\linewidth}{!}{
    \centering
    \begin{tabular}{c|c|c|c|c|c}
    \hline
    \textbf{Optimization}            & \textbf{Opt1}               & \textbf{Opt2}             & \textbf{Opt3}             & \textbf{Opt4}             & \textbf{Opt5} \\
     \hline  \hline
    Shared-buffers to FIFOs       & \textcolor{green}{\cmark}   & \textcolor{green}{\cmark} & \textcolor{green}{\cmark} & \textcolor{green}{\cmark} & \textcolor{green}{\cmark} \\
    Graph/Node-level Pipelining   & \textcolor{red}{\xmark}     & \textcolor{green}{\cmark} & \textcolor{red}{\xmark}   & \textcolor{green}{\cmark} & \textcolor{green}{\cmark} \\
    Node-level Parallelizaiton    & \textcolor{red}{\xmark}     & \textcolor{red}{\xmark}   & \textcolor{green}{\cmark} & \textcolor{green}{\cmark} & \textcolor{green}{\cmark} \\
    Combined Optimization         & \textcolor{red}{\xmark}     & \textcolor{red}{\xmark}   & \textcolor{red}{\xmark}   & \textcolor{red}{\xmark}   & \textcolor{green}{\cmark} \\
    \hline
    \end{tabular}
}
\caption{Different Levels of Optimizations of Stream-HLS}
\label{tab:opts_info}
\vspace{-0.3in}
\end{table}

%% file: tables/compare.tex
\begin{table*}[t]
\resizebox{\textwidth}{!}{
    \centering
    \begin{tabular}{c|l|l|l|l|l|l|l}
    \hline
\textbf{App.~\cite{Polybench}}        & \multicolumn{1}{c|}{\textbf{Stream-HLS (Opt5)}} & \multicolumn{2}{c|}{\textbf{Manually Optimized (Speedup$^*$)}} & \multicolumn{4}{c}{\textbf{Automatically Optimized (Speedup$^*$)}} \\ \cline{2-8}
\textbf{Medium} & \multicolumn{1}{c|}{(220, 2560, 9024) DSPs} & 
\multicolumn{1}{c|}{\textbf{Allo~\cite{allo}}} & 
\multicolumn{1}{c|}{\textbf{HeteroCL~\cite{heterocl}}}  & 
\multicolumn{1}{c|}{\textbf{HIDA~\cite{hida}}} &
\multicolumn{1}{c|}{\textbf{ScaleHLS~\cite{scalehls}}} &
\multicolumn{1}{c|}{\textbf{POM~\cite{pom}}} &
\multicolumn{1}{c}{\textbf{Vitis HLS~\cite{vitis}}} \\ \hline \hline

2mm             &3.78E+5 | 3.64E+4 | 1.18E+4 &2.68E+5 ($7.38\times$) &4.81E+5 ($13.22\times$) &2.46E+5 ($6.75\times$)  &2.46E+5 ($6.75\times$)  &1.98E+6 ($54.38\times$)  &1.18E+8 ($3236.36\times$) \\
3mm             &6.02E+5 | 4.91E+4 | 1.44E+4 &2.23E+5 ($4.54\times$) &4.65E+5 ($9.47\times$)  &2.79E+5 ($5.69\times$)  &2.66E+5 ($5.42\times$)  &2.03E+6 ($41.45\times$)  &1.28E+8 ($2600.83\times$) \\
atax            &1.59E+4 | 2.18E+3 | 2.18E+3 &8.82E+3 ($4.05\times$) &2.92E+4 ($13.39\times$) &1.34E+4 ($6.15\times$)  &1.93E+4 ($8.85\times$)  &Failed                   &1.29E+6 ($591.09\times$) \\
bicg            &8.03E+3 | 1.11E+3 | 1.11E+3 &1.65E+3 ($1.48\times$) &2.88E+3 ($2.59\times$)  &1.44E+4 ($12.99\times$) &1.44E+4 ($12.99\times$) &Failed                   &1.28E+6 ($1150.72\times$) \\
gemm            &3.30E+5 | 2.41E+4 | 7.10E+3 &2.39E+5 ($9.92\times$) &2.36E+5 ($9.81\times$)  &1.77E+5 ($7.36\times$)  &1.77E+5 ($7.36\times$)  &1.33E+6 ($55.34\times$)  &1.06E+7 ($441.14\times$) \\
gesummv         &3.18E+3 | 6.73E+2 | 6.73E+2 &1.54E+3 ($2.29\times$) &1.79E+3 ($2.66\times$)  &6.32E+3 ($9.38\times$)  &6.32E+3 ($9.40\times$)  &6.58E+3 ($9.78\times$)   &5.00E+5 ($742.97\times$) \\
mvt             &8.04E+3 | 6.67E+2 | 5.50E+2 &4.68E+2 ($0.70\times$) &2.08E+3 ($3.12\times$)  &1.01E+4 ($15.08\times$) &1.51E+4 ($22.60\times$) &4.00E+4 ($60.02\times$)  &2.56E+6 ($3838.11\times$) \\
\hline
\multicolumn{2}{l|}{Geo. Mean for 220 DSPs Limit (POM)}   &$0.35\times$            &$0.68\times$             &$0.93\times$             &$1.03\times$             &$3.74\times$              &$145.45\times$ \\
\hline
\multicolumn{2}{l|}{Geo. Mean for 2560 DSPs Limit (1 SLR)}  &$3.17\times$            &$6.20\times$             &$8.50\times$             &$9.42\times$             &$37.40\times$             &$1325.85\times$ \\
\hline
\multicolumn{2}{l|}{Geo. Mean for 9024 DSPs U280 Limit}  &$5.42\times$            &$10.62\times$            &$14.55\times$            &$16.13\times$            &$79.43\times$             &$2270.25\times$ \\
\hline

\multicolumn{8}{l}{$^*$ Speedups compared to Stream-HLS Opt5 with the 2560 DSPs limit for one super logic region (SLR)}\\

\end{tabular}
}
\caption{RTL Simulation Cycle Count Comparison with Prior Abstraction and Automation Frameworks}
\label{tab:comp}
\vspace{-0.3in}
\end{table*}

%% file: tables/scalability.tex
\begin{table}[h]
\resizebox{\linewidth}{!}{
\centering
\begin{tabular}{c|c|c|c|c|c|c}
\hline
\multirow{2}{*}{\textbf{App.}} & \multicolumn{2}{c|}{\textbf{Stream-HLS (Opt5)}} & \multicolumn{2}{c|}{\textbf{HIDA~\cite{hida}}}  & \multicolumn{2}{c}{\textbf{POM~\cite{pom}}}\\ \cline{2-7}
              & \textbf{DSE(s)} & \textbf{DSPs(\%)} & \textbf{DSE(s)} & \textbf{DSPs(\%)} & \textbf{DSE (s)} & \textbf{DSPs(\%)} \\ \hline \hline

2mm& (7, 4, 4)& (2.5, 21.5, 82.4)& 644& 10.7& 4& 0.7\\
3mm& (6, 6, 4)& (2.5, 27.8, 88.9)& 1501& 6.9& 8& 1.0\\
atax& (4, 3, 3)& (2.2, 16.6, 16.6)& 45& 1.7& - & -\\
bicg& (3, 7, 3)& (2.2, 16.6, 16.6)& 708& 1.3& - & -\\
gemm& (4, 4, 3)& (2.1, 28.3, 92.0)& 1775& 3.6& 1& 0.6\\
gesummv& (4, 3, 4)& (2.4, 10.2, 10.2)& 839& 2.0& 2& 1.1\\
mvt& (3, 4, 4)& (2.2, 25.6, 39.7)& 825& 1.8& 2& 0.5\\
\hline
\multicolumn{3}{l|}{\textbf{DSE Runtime Geo. Mean Speedup}} & $176.41\times$ & & $0.74\times$ \\
\hline
\end{tabular}
}
DSE runtime speedups are measured over the 9024 DSPs limit
\caption{DSE Runtimes in Seconds and DSP Utilization under (220, 2560, 9024) DSP Limits}
\label{tab:scalability}
\vspace{-0.3in}
\end{table}

%% file: tables/3mm.tex
\begin{table}[h]
\resizebox{\linewidth}{!}{
    \centering
    \begin{tabular}{c|c|c|c|c}
    \hline
    \multirow{2}{*}{\textbf{Node (Ops)}} & \textbf{Opt5 - 2560}  & \textbf{HIDA}        & \textbf{Opt5 - 220}    & \textbf{POM} \\\cline{2-5}
                        & \textbf{Lat. | DSPs}  & \textbf{Lat. | DSPs} & \textbf{Lat. | DSPs}   & \textbf{Lat. | DSPs}\\
     \hline  \hline
    Gemm1(1.37E+7) & 4.56E+4 | 752  & 1.14E+5 | 303   & 5.70E+5 | 63     & 8.55E+5 | 43 \\
    Gemm2(1.76E+7) & 4.39E+4 | 1001 & 1.46E+5 | 33    & 4.39E+5 | 101    & 1.10E+6 | 3 \\
    Gemm3(1.44E+7) & 4.79E+4 | 753  & 1.33E+5 | 17    & 5.99E+5 | 62     & 8.98E+5 | 3 \\
    \hline
    Total (4.56E+7) & 4.91E+4 | 2507  & 2.79E+5 | 623   &6.02E+5 | 227     & 2.03E+6 | 92 \\
    \hline
    \end{tabular}
}
\caption{\textbf{\textit{3mm}} Latency and DSP Count Breakdown}
\label{tab:3mm_perf}
\vspace{-0.2in}
\end{table}

%% file: tables/ablation.tex
\begin{table*}[t]
\centering
\begin{tabular}{c|l|l|l|l|l}
\hline
\textbf{Application} & 
\multicolumn{1}{c|}{\textbf{Opt1 (baseline)}} & 
\multicolumn{1}{c|}{\textbf{Opt2}} &
\multicolumn{1}{c|}{\textbf{Opt3}} &
\multicolumn{1}{c|}{\textbf{Opt4}} & 
\multicolumn{1}{c}{\textbf{Opt5}}  \\ \hline \hline

Autoencoder               &	2.98E+7 ($1.00\times$) &	7.44E+6 ($4.00\times$) &	1.19E+6 ($24.91\times$) &	5.06E+5 ($58.80\times$)   &	3.92E+4 ($759.47\times$)  \\
Residual MLP              &	1.84E+7 ($1.00\times$) &	4.60E+6 ($3.99\times$) &	6.75E+5 ($27.19\times$) &	3.83E+5 ($47.98\times$)   &	3.98E+4 ($461.40\times$)  \\\hline
Residual Block            &	5.79E+7 ($1.00\times$) &	7.95E+6 ($7.28\times$) &	7.25E+6 ($7.99\times$)  &	2.09E+6 ($27.70\times$)   &	2.09E+6 ($27.67\times$)   \\
DWSConv. Block            &	6.63E+6 ($1.00\times$) &	1.04E+6 ($6.37\times$) &	3.32E+5 ($19.98\times$) &	1.35E+5 ($49.26\times$)   &	1.35E+5 ($49.26\times$)   \\\hline
Feed Forward              &	6.73E+7 ($1.00\times$) &	1.68E+7 ($4.00\times$) &	1.05E+6 ($63.85\times$) &	1.31E+5 ($511.74\times$)  &	6.60E+4 ($1019.39\times$) \\
Multi-Head Self Attention &	1.05E+7 ($1.00\times$) &	2.69E+6 ($3.91\times$) &	2.68E+5 ($39.34\times$) &	3.95E+4 ($266.82\times$)  &	3.51E+4 ($299.67\times$)  \\\hline
3mm                       &	6.38E+7 ($1.00\times$) &	8.82E+6 ($7.24\times$) &	1.28E+6 ($50.00\times$) &	4.91E+4 ($1300.42\times$) &	4.91E+4 ($1300.42\times$) \\
atax                      &	1.28E+6 ($1.00\times$) &	3.19E+5 ($4.00\times$) &	3.38E+4 ($37.84\times$) &	2.18E+3 ($586.08\times$)  &	2.18E+3 ($586.08\times$)  \\\hline
7mm Balanced              &	1.15E+6 ($1.00\times$) &	2.86E+5 ($4.00\times$) &	6.66E+4 ($17.21\times$) &	5.68E+3 ($201.52\times$)  &	5.68E+3 ($201.52\times$)  \\
7mm Imbalanced            &	4.29E+6 ($1.00\times$) &	1.08E+6 ($3.96\times$) &	1.15E+5 ($37.44\times$) &	5.16E+4 ($83.17\times$)   &	1.00E+4 ($427.74\times$)  \\\hline
Geo. Mean                 &	$1.00\times$            &	$4.70\times$            &	$28.31\times$            &	$152.30\times$             &	$314.89\times$              \\

\hline
\end{tabular}
\caption{RTL Simulation Cycle Counts under Different Optimiztion Levels of Stream-HLS using 2560 DSPs Limit}
\label{tab:ablation}
\vspace{-0.3in}
\end{table*}

%% file: sections/related_work.tex
\section{Related Work}
A plethora of research has been conducted to automate and abstract the design of hardware circuits. 
Multiple studies have focused on design space exploration of pragma/directive insertion (mainly pipelining, array partitioning, and unrolling) for HLS such as~\cite{nlp-dse, autodse, harp, lorenzo, gnn-dse, comba}. However, as shown in this work, the design space of hardware designs is much larger including for example code transformation and graph-level pipelining.

Other studies proposed stand-alone HLS compilers such as the open-source LegUp~\cite{legup1, legup2} compiler and the Dynamatic~\cite{dynamatic1, dynamatic2} compiler for dynamically-scheduled circuits suitable for irregular and control-dominated workloads. Other efforts developed Python-based abstraction frameworks that are built on top of commercial HLS tools such as PyLog~\cite{pylog}, HeteroCL~\cite{heterocl}, DaCe~\cite{dace}, and Allo~\cite{allo}. Similarly, TAPA~\cite{tapa} developed an open-source C++ library on top of Vitis HLS~\cite{vitis} for streaming dataflow designs, while Dahlia~\cite{dahlia} introduced a programming language with type-checking for predictable hardware execution of affine kernels. Other projects introduced intermediate representations and compilers suitable for hardware accelerators such as Spatial~\cite{spatial}, Calyx~\cite{calyx}, HIR~\cite{hir}, and multiple other efforts in the CIRCT project~\cite{circt}. These efforts are orthogonal to Stream-HLS automation and design space exploration.

Another line of work is automatic accelerator generation. Multiple accelerator generation frameworks have been proposed for single-kernel applications such as PolySA~\cite{polysa}, AutoSA~\cite{autosa}, and SuSy~\cite{susy} for systolic arrays. POLSCA~\cite{polsca}, ScaleHLS~\cite{scalehls}, HIDA~\cite{hida}, and POM~\cite{pom} are more general MLIR-based frameworks for affine kernels, but they lack streaming capabilities.

Other streaming-related works include FBLAS~\cite{fblas}, a streaming-based HLS library for linear algebra and Fleet~\cite{fleet} for composing and parallelizing a user's RTL processing units using streams. Additionally, Darkroom~\cite{darkroom}, SODA~\cite{soda}, and Clockwork~\cite{clockwork} proposed automatic compilers for stencil/image processing kernels, which we plan to inspire from to support stencils in Stream-HLS.

%% file: sections/conclusion.tex
\section{Future Work}
The design space we discussed in this paper focuses on the scheduling and parallelization of multi-kernel applications. Stream-HLS as well as the prior works~\cite{allo, heterocl, hida, scalehls, pom} we compared it with in Table~\ref{tab:comp} assume that input and output data come from fast partitioned on-chip memories, which may not be suitable for many use cases. We are currently working on expanding our formulation and DSE to consider the limited bandwidth of DDR and HBM channels, including data reuse and on-chip/off-chip communication tradeoffs. We are also working on extending Stream-HLS to support stencils and image processing applications similar to ~\cite{soda, clockwork, darkroom}.

\section{Conclusion}
\label{sec:conclusion}
In this paper, we introduce a novel approach for transforming sequential multi-kernel applications into parallelized dataflow architectures. We propose an accurate performance model for dataflow architectures considering both shared buffer and FIFO for inter-task communication. This model is utilized to optimize application scheduling, enhancing node and graph-level pipelining and node-level parallelization. Our methodology is implemented as an automated framework called Stream-HLS, built on the MLIR infrastructure and incorporating various optimization passes. Stream-HLS designs yield considerable speedups compared to manually and automatically optimized designs of prior SOTA frameworks.

\section{Acknowledgment and Disclosure}
This work was partially supported by NSF awards CCF-1937599 and CCF-2211557, the~\href{https://cdsc.ucla.edu/partners/} {CDSC industrial partners}, the AMD\footnote{Jason Cong has a financial interest in AMD.} HACC Programs, and the PRISM Center under the JUMP2.0 Program sponsored by Semiconductor Research Corporation (SRC) and DARPA.